\documentclass[11pt]{amsart}
\usepackage{amsmath,amsfonts,amscd,amssymb,epsf}
\newtheorem{theorem}{Theorem}[section]
\newtheorem{proposition}[theorem]{Proposition}

\newtheorem{corollary}[theorem]{Corollary}
\theoremstyle{definition}

\theoremstyle{remark} \newtheorem{remark}[theorem]{Remark}
\numberwithin{equation}{section}

\DeclareMathOperator{\ad}{ad} \DeclareMathOperator{\Res}{Res}
\DeclareMathOperator{\Mobius}{M\ddot{o}bius}
\DeclareMathOperator{\fixed}{fixed}
\DeclareMathOperator{\dToda}{dToda}

\newcommand{\Z}{{\mathbb{Z}}}

\newcommand{\C}{{\mathbb{C}}}
\newcommand{\mL}{\mathcal{L}}
\newcommand{\M}{\mathcal{M}}
\newcommand{\F}{\mathcal{F}}
\newcommand{\mP}{\mathcal{P}}

\newcommand{\pa}{\partial}

\begin{document}
\title[Analytic functions and integrable hierarchies]
{Analytic functions and integrable hierarchies-characterization of tau functions}
\author{Lee-Peng Teo}
\address{Department of Applied Mathematics \\
National Chiao Tung University, 1001 \\ Ta-Hsueh Road, Hsinchu
City, 30050 \\ Taiwan, R.O.C.}  \email{lpteo@math.nctu.edu.tw}
\begin{abstract}
We prove the dispersionless Hirota equations for the
dispersionless Toda, dispersionless coupled modified KP and
dispersionless KP hierarchies using an idea from classical complex
analysis. We also prove that the Hirota equations characterize the
tau functions for each of these hierarchies. As a result, we
establish the links between the hierarchies.
\end{abstract}
 \maketitle
\section{Introduction}
Dispersionless integrable hierarchies have been under active
research in recent years (see, e.g. \cite{K2, K1, TT2, TT5, TT4,
TT1}). One of the reasons is due to its close relation with the
other area of mathematics and physics, such as topological field
theory, string theory, $2$D-gravity, matrix models and conformal
maps (see, e.g. \cite{TT3, TT6, D1, D2, BX1, BX2, BS, WZ, KKMWZ,
BMRWZ, Z, MWZ}). The tau functions of the dispersionless
integrable hierarchies play an important role in topological field
theories (\cite{D3, D4}, for they give solutions to the so-called
WDVV equation. Conversely, in \cite{BMRWZ}, it was proved that the
tau functions of the dispersionless KP (dKP) and dispersionless
Toda (dToda) hierarchies satisfy the associativity equation. One
of the main ingredients of the proof of \cite{BMRWZ} is the
dispersionless Hirota equations satisfied by the dKP and dToda
hierarchies. The dispersionless Hirota equation for the dKP
hierarchy was first derived by Takasaki and Takebe \cite{TT1} as
the dispersionless limit of the differential Fay identity. Later,
the Hirota equation was further studied by Carroll and Kodama
\cite{CK}. In connection to conformal mappings which give rise to
solutions of the dToda hierarchy, Wiegmann, Zabrodin et al derive
the Hirota equations for the dToda hierarchy \cite{WZ, MWZ,
KKMWZ}. In this paper, we point out that these Hirota equations
are closely related to some concepts in classical complex
analysis, namely Faber polynomials and Grunsky coefficients.

In Section 2, we review some concepts from classical complex
analysis. We define some classes of formal power series which
appear in dKP, dcmKP (dispersionless coupled modified KP , see
\cite{T}) and dToda hierarchies. We generalize the definition of
the Grunsky coefficients and Faber polynomials to these classes of
formal power series and review their properties. In Section 3, we
review the dToda, dcmKP and dKP hierarchies and their tau
functions. We derive the dispersionless Hirota equations by
establishing the relation between the tau functions and the
Grunsky coefficients. We also prove that the dispersionless Hirota
equations for each of these hierarchies uniquely characterize the
tau functions of their solutions. As a corollary, we show that
some solutions of the dToda hierarchy will give rise to solutions
of dcmKP hierarchy, which in turn will give rise to solutions of
dKP hierarchy.

\section{Algebraic analysis}
\subsection{Spaces of formal power series}
 We consider the following classes
of formal power series :

\begin{align*}
\tilde{\Sigma}& =\left\{ g(z) = b z + b_0 +
\frac{b_1}{z}+\ldots =bz+\sum_{n=0} ^{\infty} b_n z^{-n}; b \neq 0\right\},\\
\Sigma &=\left\{ g(z) =  z + b_0 + \frac{b_1}{z}+\ldots=
z+\sum_{n=0}^{\infty} b_n z^{-n} \right\},\\
\Sigma_0 &=\left\{ g(z) =  z  +
\frac{b_1}{z}+\frac{b_{2}}{z^2}+\ldots= z+\sum_{n=1}^{\infty} b_n
z^{-n} \right\}.
\end{align*}

$\tilde{\Sigma}$ can be considered as completion \footnote{The
completion is with respect to the filtration $\tilde{\Sigma}_N=
\left\{g(z)=  bz + \sum_{n=N}^{\infty} b_n z^{-n}, b\neq 0
\right\}$ on $\tilde{\Sigma}$.} of the space of analytic functions
that fix the point $\infty$ and univalent in a small neighbourhood
of $\infty$. $\Sigma$ and $\Sigma_0$ are subspaces of
$\tilde{\Sigma}$ consist of formal power series satisfying certain
normalization conditions. By post-composing $\tilde{g} \in
\tilde{\Sigma}$ with the linear map $z \mapsto (1/\tilde{b})z$, we
get a function $g$ in $\Sigma$. Further post-composition with the
linear map $z \mapsto z-b_0$, we get a function $g_0 $ in
$\Sigma_0$. As their counterpart, we consider another three
classes of formal power series:

\begin{align*}
\tilde{S} &= \left\{f(z) = a_1 z+ a_2 z^2 + a_3 z^3 +
\ldots=\sum_{n=1}^{\infty} a_n z^n; a_1\neq 0 \right\},\\
S &= \left\{f(z) =  z+ a_2 z^2 + a_3 z^3 + \ldots=z+
\sum_{n=2}^{\infty} a_n z^n\right\},\\
S_0 &= \left\{f(z) =  z+ a_3 z^3 +a_4 z^4+ \ldots=z+
\sum_{n=3}^{\infty} a_n z^n\right\}.
\end{align*}

$\tilde{S}$ can be considered as completion of the space of
analytic functions fixing the point $0$ and univalent in a
neighbourhood of $0$. $S$ and $S_0$ are subspaces of $\tilde{S}$
consist of formal power series subjecting to additional
normalization conditions. By post-composing $ \tilde{f} \in
\tilde{S}$ with the linear map $z \mapsto (1/\tilde{a}_1) z$, we
get a function $f \in S$. Further post-composition with the
$\Mobius$ transformation $z\mapsto z/(1+a_2z)$, we get a function
$f_0 \in S_0$. Observe that the map \[ f \mapsto g(z)=
\frac{1}{f(\frac{1}{z})}
\]
is a bijection between $\tilde{S}$ and $\tilde{\Sigma}$, $S$ and
$\Sigma$, $S_0$ and $\Sigma_0$ respectively.

\subsection{Faber polynomials and Grunsky coefficients}
We review and generalize some concepts from classical complex
analysis. For details, see \cite{Pom, Duren}.

One of the important problems in classical complex analysis is the
determination of the upper bound satisfied by the coefficients of
$a_n$'s in order that $f \in S$ is univalent on the unit disc. A
lot of effort has been devoted to the proof of the famous
Bieberbach conjecture (1916) : If $f \in S$ is univalent on the
unit disc, then $|a_n| \leq n$ \footnote{This conjecture was
completely proved by de Branges in 1984 \cite{deB}.}. In the early
attempts of the proof, one of the important tools is the Grunsky's
inequality and its generalization. In 1939, Grunsky found a
sequence of inequalities that should be satisfied by the so called
Grunsky coefficients in order that $g \in \Sigma$ is univalent on
$\{z \;\;\bigr\vert \;\;|z|>1\}$. Surprisingly, we found this
Grunsky coefficients appear everywhere in dispersionless limit of
integrable hierarchies (especially in association with tau
functions) without being realized its connection to complex
analysis. This is the purpose of this paper to point out this
connection, and thus give simplified proofs of some of the facts
related to dispersionless integrable hierarchies.

 First, we introduce the Faber polynomials. For $g\in \tilde{\Sigma}$ that is
analytic in a neighbourhood  of $\infty$ and $w \in \C$, consider
the function $\log \frac{g(z) -w}{bz}$. It defines an analytic
function for large $|z|$ and vanishes at $\infty$. Hence it has an
expansion at $\infty$ which can be written as
\begin{align}\label{Faber1}
\log \frac{g(z) -w}{bz} = -\sum_{n=1}^{\infty} \frac{\Phi_n(w)}{n}
z^{-n}.
\end{align}
$\Phi_n$ is called the $n$-th Faber polynomial of $g$.
Differentiate \eqref{Faber1} with respect to $z$ and define
$\Phi_0(w) \equiv 1$, we have
\begin{align} \label{Faber2}
\frac{ g'(z)}{g(z) -w}= \sum_{n=0}^{\infty} \Phi_n(w) z^{-n-1}.
\end{align}
Putting $g(z) = b z + b_0 + \sum_{n=1}^{\infty} b_n z^{-n}$ into
this expression, we have
\begin{align*}
 b - \sum_{n=1}^{\infty}nb_n z^{-n-1} = \left( b z + b_0
 -w + \sum_{n=1}^{\infty} b_n z^{-n} \right) \left( \frac{1}{z}+
 \sum_{n=1}^{\infty} \Phi_n z^{-n-1}\right).
\end{align*}
Comparing coefficients, we have
\[
\Phi_1(w) =\frac{w-b_0}{b}
\]
and the recursion formula
\[
\Phi_{n+1}(w) = \frac{w-b_0}{b} \Phi_{n} (w) - \frac{1}{b}
\sum_{k=1}^{n-1} b_{n-k} \Phi_k(w) - (n+1)\frac{b_n}{b},
\]
which can be used to solve for $\Phi_n$. From this, we see that
$\Phi_n(w)$ is a polynomial of degree $n$ of the form
\begin{align*}
\Phi_n(w) =\left(\frac{w-b_0}{b}\right)^n -n\frac{b_1}{b}
\left(\frac{w-b_0}{b}\right)^{n-2}+ \ldots.
\end{align*}
 The first few are given by
\begin{align*}
\Phi_2(w) &= \left(\frac{w-b_0}{b}\right)^2 - 2 \frac{b_1}{b}, \\
\Phi_3(w) &= \left(\frac{w-b_0}{b}\right)^3 - 3
\frac{b_1}{b}\frac{w-b_0}{b}-3\frac{b_2}{b},\\
\Phi_4(w) &= \left(\frac{w-b_0}{b}\right)^4 - 4
\frac{b_1}{b}\left(\frac{w-b_0}{b}\right)^2
-4\frac{b_2}{b}\frac{w-b_0}{b}+
2\left(\frac{b_1}{b}\right)^2-4\frac{b_3}{b}.
\end{align*}

Now we introduce the Grunsky coefficients. If $g \in
\tilde{\Sigma}$ is univalent in a neighbourhood $U$ of $\infty$,
the function $\log \frac{g(z) - g(\zeta)}{z -\zeta}$ is analytic
in the neighbourhood $U \times U$ of $(\infty, \infty)$. Its
expansion about $(\infty, \infty)$ has the form
\begin{align}\label{Grunsky1}
\log \frac{g(z) - g(\zeta)}{z -\zeta} = \log b
-\sum_{m=1}^{\infty} \sum_{n=1}^{\infty} b_{mn} z^{-m} \zeta^{-n}.
\end{align}
$b_{mn}$'s are known as Grunsky coefficients of $g$. They are
symmetric, i.e. $b_{mn} = b_{nm}$. If we put $w =g(\zeta)$ into
\eqref{Faber1} , we have
\begin{align*}
\log \frac{g(z)-g(\zeta)}{z-\zeta} = \log b -\sum_{n=1}^{\infty}
\frac{1}{n} \left(\Phi_n(g(\zeta)) - \zeta^n\right) z^{-n}.
\end{align*}
Compare with \eqref{Grunsky1}, we have
\begin{align}\label{Faber3}
\Phi_n (g(\zeta)) = \zeta^n + n\sum_{m=1}^{\infty} b_{nm}
\zeta^{-m}.
\end{align}

In fact, the Faber polynomials can be uniquely characterized as
follows: If $P_n(w)$ is a degree $n$ polynomial such that
$P_n(g(\zeta)) - \zeta^n \rightarrow 0$ as $\zeta \rightarrow
\infty$, then $P_n = \Phi_n$.

Since $\Phi_1(w) =(w-b_0)/b$, the $n=1$ case of \eqref{Faber3}
gives
\[
g(\zeta) =b\zeta+ b_0 + b\sum_{m=1}^{\infty} b_{1m} \zeta^{-m}.
\]
Hence $b_{1m}=b_{m1} = b_m/b$. In fact, using the recursion
formula for $\Phi_n$ and \eqref{Faber3}, we can recursively solve
for $b_{nm}$'s. The first fews are given by
\begin{align*}
b_{22} = \frac{b_3}{b}+ \frac{1}{2}\left(\frac{b_1}{b}\right)^2,
\hspace{0.5cm} b_{23}= \frac{b_4}{b}+ \frac{b_1b_2}{b^2},
\hspace{0.5cm} b_{33} = \frac{b_5}{b} + \frac{b_1b_3}{b^2} +\left(
\frac{b_2}{b}\right)^2 + \frac{1}{3} \left(\frac{b_1}{b}\right)^3.
\end{align*}

There is another characterization of the Faber polynomials which
plays an important role in our discussion later. Let $z=G(w)= w/b
+ \sum_{n=0}^{\infty} c_n w^{-n}$ be the inverse function of
$w=g(z)$ in the neighbourhood $U$ where $g$ is univalent. The
coefficients $c_n$ can be solved recursively. $c_0 = -b_0/b$ and
for $n \geq 1$, $c_n$ is a polynomial in $b, b_0, b_1, \ldots,
b_n$
. Let $C$ be a simple closed curve winding negatively about the
point $\infty$, lying entirely in $U$. Then $C' = g(C)$ is also a
simple closed curve winding negatively about $\infty$. Observe
that the line integral of a function $(1/2\pi i)f$ along $C$ (or
$C'$) amounts to taking residue of $f$ with respect to $\infty$.
Hence from \eqref{Faber2}, we have
\begin{align*}
\Phi_n(w) = \frac{1}{2\pi i} \oint_C \frac{g'(z)z^n}{g(z) - w}dz.
\end{align*}
Making a change of variable $z=G(\zeta)$ or equivalently
$\zeta=g(z)$, we have
\begin{align*}
\Phi_n(w) = \frac{1}{2\pi i} \oint_{C'} \frac{G(\zeta)^n}{\zeta
-w} d\zeta.
\end{align*}
Using the expansion about $\zeta = \infty$, $G(\zeta)^n =
\sum_{m=-\infty}^{n} c_{n,m} \zeta^m$ and $1/(\zeta-w)=
\sum_{k=0}^{\infty} w^k \zeta^{-k-1}$, we obtain immediately
$\Phi_n(w) = \sum_{m=0}^{n} c_{n,m} w^m$. Namely, $\Phi_n(w)$ is
the polynomial part of $G(w)^n$, which we denote by
$(G(w)^n)_{\geq 0}$, i.e.
\begin{align}\label{positive1}
\Phi_n(w) = (G(w)^n)_{\geq 0}.
\end{align}
 In general, if $A= \sum_{n =-\infty}^{\infty}
A_n w^n$ is a formal power series, and $S$ a subset of integers,
we define $(A)_{S}=\sum_{n\in S} A_n w^n$.

The Grunsky coefficients are generalized to a pair of functions
$f$ and $g$ as follows. Let $f \in \tilde{S}$ be univalent in a
neighbourhood $V$ of $0$ and $g \in \Sigma$ be univalent in a
neighbourhood $U$ of $\infty$. We say that $(f, g)$ are disjoint
relative to $(U, V)$ if the sets $f(V)$ and $g(U)$ are disjoint.
In this case, the functions
\begin{align*}
\log \frac{g(z)-g(\zeta)}{z-\zeta}, \hspace{1cm}
\log\frac{g(z)-f(\zeta)}{z}, \hspace{1cm}
\log\frac{f(z)-f(\zeta)}{z-\zeta}
\end{align*}
are analytic in $U\times U$, $U\times V$ and $V\times V$
respectively. Hence we can write down their series expansion about
$(\infty, \infty)$, $(\infty, 0)$ and $(0,0)$ respectively:
\begin{align}
\log \frac{g(z)-g(\zeta)}{z-\zeta}=
-\sum_{m=1}^{\infty}\sum_{n=1}^{\infty} b_{mn} z^{-m}\zeta^{-n},\label{Grunsky2}\\
\log \frac{g(z) - f(\zeta)}{z} = -\sum_{m=1}^{\infty}
\sum_{n=0}^{\infty} b_{m,-n} z^{-m} \zeta^n,\label{Grunsky3}\\
\log \frac{f(z)-f(\zeta)}{z-\zeta} = -\sum_{m=0}^{\infty}
\sum_{n=0}^{\infty} b_{-m, -n} z^{m} \zeta^n.\label{Grunsky4}
\end{align}
Obviously, when $m,n$ are both positive or both negative, $b_{mn}
=b_{nm}$. Hence for $m\geq 0$, $n>0$, we define $b_{-m, n}=b_{n,
-m}$. Letting $\zeta=0$ in \eqref{Grunsky3} and \eqref{Grunsky4},
we obtain
\begin{align}\label{Grunsky5}
\log\frac{g(z)}{z} = -\sum_{m=1}^{\infty} b_{m,0}
z^{-m},\hspace{1cm} \log\frac{f(z)}{z} =-\sum_{m=0}^{\infty}
b_{-m, 0} z^{m}.
\end{align}
In particular, $b_{00} = -\log a_1$. We define the generalized
Faber polynomials $\Psi_n(w)$ for $f$ by
\begin{align}\label{Faber4}
\log \frac{w- f(z)}{w}= \log \frac{f(z)}{a_1
z}-\sum_{n=1}^{\infty} \frac{\Psi_n(w)}{n} z^{n}.
\end{align}

To see the relations between $\Psi_n$ and the Grunsky
coefficients, we define the function $g_f \in \tilde{\Sigma}$ by
\[
g_f(z) = \frac{1}{f(\frac{1}{z})}= \frac{z}{a_1}
-\frac{a_2}{a_1^2}+\ldots.
\]
Using the second equation in \eqref{Grunsky5}, equations
\eqref{Grunsky3}, \eqref{Grunsky4} and \eqref{Faber4} can be
rewritten in terms of $g_f$:
\begin{align}
\log \left(1-\frac{1}{g(z) g_f(\zeta)}\right) &=
-\sum_{m=1}^{\infty}\sum_{n=1}^{\infty} b_{m,-n} z^{-m} \zeta^{-n}, \label{Grunsky7}\\
 \log \frac{g_f(z) -g_f(\zeta)}{z-\zeta} &= -\log a_1
-\sum_{m=1}^{\infty} \sum_{n=1}^{\infty} b_{-m,-n}
z^{-m}\zeta^{-n},\label{Grunsky8}\\
\log \frac{g_f(z)-\frac{1}{w}}{z} &=-\log a_1 -\sum_{n=1}^{\infty}
\frac{\Psi_n(w)}{n} z^{-n}.\label{Grunsky6}
\end{align}
Hence, $b_{-m,-n}$'s are Grunsky coefficients of $g_f$ and $\Psi_n
(w)$ is a polynomial of degree $n$ in $1/w$. If we denote by
$z=G_f(w)$ the inverse of $w=g_f(z)$, and by $z=F(w)$ the inverse
of $w=f(z)$, then
\[
F(w) =\frac{1}{G_f(\frac{1}{w})}.
\]
Equation \eqref{positive1} implies that
\[
\Psi_n(w^{-1}) = (G_f(w)^n)_{\geq 0}\] or equivalently
\begin{align}\label{Faber5}
\Psi_n(w) = (G_f(w^{-1})^n)_{\leq 0}=(F(w)^{-n})_{\leq 0}.
\end{align}
Now we derive the counterparts of \eqref{Faber3}. First, compare
\eqref{Grunsky6} to \eqref{Faber1} and \eqref{Faber3}, we have
\begin{align*}
\Psi_n (g_f(\zeta)^{-1})= \zeta^n + n\sum_{m=1}^{\infty} b_{-n,-m}
\zeta^{-m}
\end{align*}
or equivalently
\begin{align}\label{Faber7}
\Psi_n (f(\zeta))= \zeta^{-n} + n\sum_{m=1}^{\infty} b_{-n,-m}
\zeta^{m}.
\end{align}
Next, we put $w=f(\zeta)$ into \eqref{Faber1} and compare with
\eqref{Grunsky3}, we obtain
\begin{align}\label{Faber8}
\Phi_n (f(\zeta)) = n\sum_{m=0}^{\infty} b_{n, -m} \zeta^m.
\end{align}
Finally, putting $w=g(\zeta)$ into \eqref{Faber4}, we have
\begin{align*}
\log \frac{g(\zeta) - f(z)}{\zeta} = \log \frac{g(\zeta)}{\zeta} +
\log \frac{f(z)}{a_1 z} -\sum_{n=1}^{\infty}
\frac{\Psi_n(g(\zeta))}{n} z^{n}.
\end{align*}
Compare with \eqref{Grunsky3} and using equations in
\eqref{Grunsky5}, we have
\begin{align}\label{Faber9}
\Psi_n(g(\zeta)) = -nb_{-n,0} + n\sum_{m=1}^{\infty} b_{m,-n}
\zeta^{-m}.
\end{align}

For the convenience of next section, we gather again the formulas
\eqref{Grunsky5}, and the formulas of the Faber polynomials in
terms of the Grunsky coefficients
\eqref{Faber3},\eqref{Faber7},\eqref{Faber8}, \eqref{Faber9}.
\begin{align}\label{Faber10}
\log\frac{g(z)}{z} &= -\sum_{m=1}^{\infty} b_{m,0}
z^{-m},\hspace{1.5cm} \log\frac{f(z)}{z} =-\sum_{m=0}^{\infty}
b_{-m, 0} z^{m}\\
\Phi_n (g(\zeta)) &= \zeta^n + n\sum_{m=1}^{\infty} b_{nm}
\zeta^{-m},\hspace{1cm}\Phi_n (f(\zeta)) =
nb_{n,0}+n\sum_{m=1}^{\infty}
b_{n, -m} \zeta^m, \nonumber\\
\Psi_n(g(\zeta)) &= -nb_{-n,0} + n\sum_{m=1}^{\infty} b_{m,-n}
\zeta^{-m},\hspace{0.4cm}\Psi_n (f(\zeta))= \zeta^{-n} +
n\sum_{m=1}^{\infty} b_{-n,-m} \zeta^{m}.\nonumber
\end{align}

The analysis above can be extended formally to the whole space
$\tilde{\Sigma}$ and $\tilde{S}$. All the Taylor (Laurent)
expansions are considered as formal power series expansions. All
the identities hold formally.

\section{Dispersionless hierarchies and tau functions}
We quickly review dispersionless Toda (dToda), dispersionless
coupled modified KP (dcmKP) and dispersionless KP (dKP)
hierarchies and their tau functions. For details, see \cite{TT2,
TT4, TT1, T}. For each of these dispersionless hierarchies, we
give a new derivation of the dispersionless Hirota equation
satisfied by the tau function, using the algebraic analysis we
discussed in the previous section. We also prove that the
dispersionless Hirota equations uniquely characterize the tau
functions.
\subsection{Dispersionless Toda hierarchy}

\subsubsection{The hierarchy}
The fundamental quantities in dToda hierarchy are two formal power
series in $p$:
\begin{align*}
\mL(p)=& p + \sum_{n=0}^{\infty} u_{n+1}(t) p^{-n},\\
\tilde{\mL}^{-1}(p) =& \tilde{u}_0(t)p^{-1} + \sum_{n=0}^{\infty}
\tilde{u}_{n+1}(t) p^n.
\end{align*}
Here $u_n(t)$ and $\tilde{u}_n(t)$ are functions of the
independent variables $t_n, n \in \Z$, which we denote
collectively by $t$. The Lax representation is \footnote{Here it
is understood that $p$ is a formal variable and does not depend on
$t$.}
\begin{align}\label{Lax1}
\frac{\pa \mL}{\pa t_n} =\{ (\mL^n)_{\geq 0}, \mL\}_T,
\hspace{2cm}
\frac{\pa \mL}{\pa t_{-n}} = \{(\tilde{\mL}^{-n})_{<0}, \mL\}_T,\\
\frac{\pa \tilde{\mL}}{\pa t_n} =\{ (\mL^n)_{\geq 0},
\tilde{\mL}\}_T, \hspace{2cm} \frac{\pa \tilde{\mL}}{\pa t_{-n}} =
\{(\tilde{\mL}^{-n})_{<0}, \tilde{\mL}\}_T.\nonumber
\end{align}
Here $\{ \cdot, \cdot\}_T$ is the Poisson bracket for dToda
hierarchy
\[
\{f, g\}_T =p \frac{\pa f}{\pa p} \frac{\pa g}{\pa t_0}-p\frac{\pa
f}{\pa t_0} \frac{\pa g}{\pa p}.
\]
There exists a function $\phi(t)$ such that
\begin{align}\label{phi1}
\frac{\pa \phi}{\pa t_0} = \log \tilde{u}_0, \hspace{1cm}
\frac{\pa \phi}{\pa t_n} = (\mL^n)_{0}, \hspace{1cm} \frac{\pa
\phi}{\pa t_{-n}} = -(\tilde{\mL}^{-n})_0
\end{align}
and two functions $\varphi = \sum_{n=1}^{\infty} \varphi_n
p^{-n}$, $\psi=\sum_{n=1}^{\infty} \psi_n p^n$ such that
\begin{align*}
\mL =& e^{\ad \varphi} p, \hspace{3.5cm} \tilde{\mL} = e^{\ad
\tilde{\varphi}} p,\\
\nabla_{t_n, \varphi} \varphi =& -(\mL^n)_{<0}, \hspace{2cm}
\nabla_{t_n, \tilde{\varphi}}\tilde{\varphi}= (\mL^n)_{\geq 0},\\
\nabla_{t_{-n}, \varphi} \varphi =& (\tilde{\mL}^{-n})_{<0}
\hspace{2.2cm} \nabla_{t_{-n}, \tilde{\varphi}}\tilde{\varphi}=-
(\tilde{\mL}^{-n})_{\geq 0}.
\end{align*}
Here $(\ad f)(g) = \{f, g\}_T$, $\tilde{\varphi}$ is defined such
that $e^{\ad \tilde{\varphi}} =e^{\ad \phi} e^{\ad \psi}$ and
\[
\nabla_{t, A} B =\sum_{n=0}^{\infty} \frac{(\ad A)^n}{(n+1)!}
\frac{\pa B}{\pa t}.
\]
The Orlov-Schulman functions are defined by
\begin{align}\label{Orlov1}
\M &= e^{\ad \varphi}\left(\sum_{n=1}^{\infty} nt_n p^{n}
+t_0\right) = \sum_{n=1}^{\infty} nt_n \mL^n + t_0+
\sum_{n=1}^{\infty} v^T_n
\mL^{-n},\\
\tilde{\M} &= e^{\ad \tilde{\varphi}}\left(-\sum_{n=1}^{\infty}
nt_{-n} p^{-n} +t_0\right) = -\sum_{n=1}^{\infty} nt_{-n}
\tilde{\mL}^{-n} + t_0 - \sum_{n=1}^{\infty} \tilde{v}^T_n
\tilde{\mL}^{n}. \nonumber
\end{align}
There exists a tau function $\tau_{\dToda}$ such that
\begin{align*}
\frac{\pa \log \tau_{\dToda}}{\pa t_0} = \phi,
\hspace{1cm}\frac{\pa \log \tau_{\dToda}}{\pa t_n} =v^T_n
,\hspace{1cm}\frac{\pa \log \tau_{\dToda}}{\pa
t_{-n}}=\tilde{v}^T_n.
\end{align*}
The tau function appears to be the dispersionless limit (leading
term) of the tau function of a corresponding Toda hierarchy. The
free energy of the hierarchy is defined by $\F= \log
\tau_{\dToda}$.
\subsubsection{Dispersionless Hirota equation} The tau
function generates the coefficients of $\mL$. More precisely, we
have the following identities:
\begin{align}\label{identity3}
\log p &=\log \mL -\sum_{m=1}^{\infty} \frac{1}{m}\frac{\pa^2 \F
}{\pa t_0 \pa t_m}\mL^{-m},\\
 (\mL^n)_{\geq 0} &= \mL^n -
\sum_{m=1}^{\infty}\frac{1}{m} \frac{\pa^2 \F}{\pa t_n \pa t_m}
\mL^{-m} =\frac{\pa^2 \F}{\pa t_0\pa t_n} - \sum_{m=1}^{\infty}
\frac{1}{m}\frac{\pa^2 \F}{\pa t_{-m}\pa t_n}
\tilde{\mL}^m,\nonumber\\
\log p &=\log \tilde{\mL} + \frac{\pa^2 \F}{\pa
t_0^2}-\sum_{m=1}^{\infty} \frac{1}{m}
\frac{\pa^2\F}{\pa t_{-m}\pa t_0} \tilde{\mL}^m, \nonumber\\
(\tilde{\mL}^{-n})_{<0} &=-\sum_{m=1}^{\infty} \frac{1}{m}
\frac{\pa^2 \F}{\pa t_m \pa t_{-n}} \mL^{-m} = \tilde{\mL}^{-n}
+\frac{\pa^2 \F}{\pa t_0\pa t_{-n}}-\sum_{m=1}^{\infty}
\frac{1}{m} \frac{\pa^2 \F}{\pa t_{-m} \pa
t_{-n}}\tilde{\mL}^{m}\nonumber.
\end{align}

Now if we identify $p$ with $w$, $\mL$ with $z$, then the first
equation define $w$ as a function of $z$. Obviously, it belongs to
the space $\Sigma$. We identify this function with our $w=g(z)$ in
Section 2. In other words, $\mL(p)$ is identified with the
function $z=G(w)$, the inverse of $g$. Similarly, the third
equation defines $w$ as a function of $\tilde{\mL}$, which belongs
to $\tilde{S}$. We identify this function with $w=f(z)$ or
equivalently, $\tilde{\mL}(p)$ is identified with $z=F(w)$, the
inverse of $f$.
 Under these identifications, we see that the Faber polynomials
 $\Phi_n(w)$'s are identified with $(\mL^n(w))_{\geq 0}$, and
 $\Psi_n(w)$'s are identified with
 $(\tilde{\mL}^{-n}(w))_{\leq 0}$. Now compare \eqref{identity3}
 with \eqref{Faber10}, we find that the Grunsky coefficients $b_{nm}$ of
 the pair $(g=w(\mL), f=w(\tilde{\mL}))$ are related to the tau
 function or free energy by
 \begin{align}\label{Grunsky9}
b_{00} &= -\frac{\pa^2 \F}{\pa t_0^2}, \hspace{1cm}b_{n,0}
=\frac{1}{n} \frac{\pa^2 \F}{\pa t_0 \pa t_n},\hspace{1cm}
b_{-n,0} = \frac{1}{n} \frac{\pa^2 \F}{\pa t_0 \pa t_{-n}},
\hspace{0.5cm}n\geq 1,\\
b_{m,n} &= -\frac{1}{mn}\frac{\pa^2 \F}{\pa t_m\pa t_n} ,
\hspace{2cm} b_{-m,-n} =-\frac{1}{mn} \frac{\pa^2\F}{\pa t_{-m}
\pa t_{-n}},
\hspace{0.5cm}n,m\geq 1, \nonumber \\
b_{-m,n}&=b_{n,-m}=-\frac{1}{mn} \frac{\pa^2 \F}{\pa t_{-m}
t_n},\hspace{0.5cm}n,m\geq 1.\nonumber
 \end{align}
 From \eqref{identity3}, we can express $f$ and $g$ in terms of the tau function or free
 energy:
 \begin{align*}
g(z) = z\exp\left(-\sum_{m=1}^{\infty} \frac{1}{m}\frac{\pa^2 \F
}{\pa t_0 \pa t_m}z^{-m}\right) ,\hspace{0.4cm} f(z) =z \exp\left(
\frac{\pa^2 \F}{\pa t_0^2}-\sum_{m=1}^{\infty} \frac{1}{m}
\frac{\pa^2\F}{\pa t_{-m}\pa t_0} z^m \right).
 \end{align*}
 As before, we define $g_f(z) =1/ f(z^{-1})$, then
 \begin{align*}
g_f(z) = z\exp\left(- \frac{\pa^2 \F}{\pa
t_0^2}+\sum_{m=1}^{\infty} \frac{1}{m} \frac{\pa^2\F}{\pa
t_{-m}\pa t_0} z^{-m} \right).
 \end{align*}
 Hence rewriting the definition of the generalized Grunsky
 coefficients in terms of the tau function or free energy, we
 obtain the Hirota equation for dispersionless Toda
 hierarchy. Namely, from \eqref{Grunsky2}, \eqref{Grunsky7},
 \eqref{Grunsky8}, we have
\begin{align}\label{bilinear1}
&z_1\exp\left(-\sum_{m=1}^{\infty} \frac{1}{m}\frac{\pa^2 \F }{\pa
t_0 \pa t_m}z_1^{-m}\right)-z_2\exp\left(-\sum_{m=1}^{\infty}
\frac{1}{m}\frac{\pa^2 \F
}{\pa t_0 \pa t_m}z_2^{-m}\right)\\
=&(z_1-z_2)\exp\left(\sum_{m,n=1}^{\infty}
\frac{1}{mn} \frac{\pa^2 \F}{\pa t_m \pa t_n}z_1^{-m}z_2^{-n}\right),\nonumber\\
&1-\frac{1}{z_1z_2}\exp\left(\frac{\pa^2 \F}{\pa
t_0^2}+\sum_{m=1}^{\infty} \frac{1}{m}\frac{\pa^2 \F }{\pa t_0 \pa
t_m}z_1^{-m}-\sum_{m=1}^{\infty} \frac{1}{m} \frac{\pa^2\F}{\pa
t_{-m}\pa
t_0}z_2^{-m} \right)\nonumber\\
=&  \exp\left(\sum_{m,n=1}^{\infty} \frac{1}{mn} \frac{\pa^2
\F}{\pa t_m \pa t_{-n}}z_1^{-m}z_2^{-n} \right),\nonumber\\
&z_1 \exp\left( \sum_{m=1}^{\infty} \frac{1}{m} \frac{\pa^2\F}{\pa
t_{-m}\pa t_0} z_1^{-m}\right)-z_2 \exp\left( \sum_{m=1}^{\infty}
\frac{1}{m} \frac{\pa^2\F}{\pa t_{-m}\pa
t_0} z_2^{-m} \right)\nonumber\\
=&(z_1-z_2)\exp\left(\sum_{m,n=1}^{\infty} \frac{1}{mn}
\frac{\pa^2 \F}{\pa t_{-m} \pa
t_{-n}}z_1^{-m}z_2^{-n}\right).\nonumber
\end{align}
We should understand these identities as defining a sequence of
relations satisfied by the second derivatives of $\F$ by comparing
the coefficients of $z_1^{-m} z_2^{-n}$ on both sides.

Conversely, the tau function is uniquely characterized by these
Hirota equations.
\begin{proposition}\label{char1}
If $\F=\log \tau$ is a function of $t_n$, $n\in \Z$ that satisfies
the Hirota equations \eqref{bilinear1}, then $\tau$ is a tau
function of a solution of the dToda hierarchy. More explicitly, if
we define $\mL(p)$ and $\tilde{\mL}(p)$ by formally inverting the
functions $\mathfrak{p}(\mL)$ and
$\tilde{\mathfrak{p}}(\tilde{\mL})$ defined by
\begin{align}\label{identity4}
\log \mathfrak{p}(\mL) &= \log \mL-\sum_{m=1}^{\infty}
\frac{1}{m}\frac{\pa^2 \F
}{\pa t_0 \pa t_m}\mL^{-m},\\
\log \tilde{\mathfrak{p}}(\tilde{\mL}) &=\log \tilde{\mL} +
\frac{\pa^2 \F}{\pa t_0^2}-\sum_{m=1}^{\infty} \frac{1}{m}
\frac{\pa^2\F}{\pa t_{-m}\pa t_0} \tilde{\mL}^m, \nonumber
\end{align}
then $(\mL, \tilde{\mL})$ satisfies the Lax equations \eqref{Lax1}
\footnote{The coefficients of the functions $\mathfrak{p}$ and
$\tilde{\mathfrak{p}}$ depend on $t$ since they can be expressed
as functions of second derivatives of $\F$ by exponentiating
\eqref{identity4}. The dependence of the coefficients of
$\mathcal{L}$ and $\tilde{\mL}$ on $t$ are understood in the
following way: They are inverses of $\mathfrak{p}$ and
$\tilde{\mathfrak{p}}$ respectively. Their coefficients can be
solved in terms of the coefficients of $\mathfrak{p}$ and
$\tilde{\mathfrak{p}}$ and are hence functions of $t$.} for dToda
hierarchy.
\end{proposition}
\begin{proof}
 In the first part of the proof, we trace the reasoning above backward. We
define the function $g \in \Sigma$ and $f \in \tilde{S}$ by
\begin{align*}
\log \frac{g(z)}{z} =-\sum_{m=1}^{\infty} \frac{1}{m}\frac{\pa^2
\F }{\pa t_0 \pa t_m}z^{-m},\hspace{1cm} \log
\frac{f(z)}{z}=\frac{\pa^2 \F}{\pa t_0^2}-\sum_{m=1}^{\infty}
\frac{1}{m} \frac{\pa^2\F}{\pa t_{-m}\pa t_0} z^m.
\end{align*}
We define the generalized Grunsky coefficients of the pair $(f,g)$
by the equations \eqref{Grunsky2}, \eqref{Grunsky3},
\eqref{Grunsky4}. We also define $g_f(z) = 1/f(z^{-1})$, then
\[
\log \frac{g_f(z)}{z}=-\frac{\pa^2 \F}{\pa
t_0^2}+\sum_{m=1}^{\infty} \frac{1}{m} \frac{\pa^2\F}{\pa
t_{-m}\pa t_0}z^{-m}.
\]
In terms of $g$ and $g_f$, the Hirota equations \eqref{bilinear1}
read as
\begin{align*}
\log \frac{g(z_1)-g(z_2)}{z_1 - z_2} &=\sum_{m,n=1}^{\infty}
\frac{1}{mn} \frac{\pa^2 \F}{\pa t_m \pa t_n}z_1^{-m}z_2^{-n},\\
\log\left(1-\frac{1}{g(z_1) g_f(z_2)}\right)
&=\sum_{m,n=1}^{\infty} \frac{1}{mn} \frac{\pa^2 \F}{\pa t_m \pa
t_{-n}}z_1^{-m}z_2^{-n},\\
\log \frac{g_f(z_1) - g_f(z_2)}{z_1 -z_2} &=-\frac{\pa^2\F}{\pa
t_0^2}+\sum_{m,n=1}^{\infty} \frac{1}{mn} \frac{\pa^2 \F}{\pa
t_{-m} \pa t_{-n}}z_1^{-m}z_2^{-n}.
\end{align*}
Comparing with \eqref{Grunsky2}, \eqref{Grunsky5},
\eqref{Grunsky7}, \eqref{Grunsky8}, we find the Grunsky
coefficients in terms of $\F$ are given by the equations in
\eqref{Grunsky9}. Hence if we define $\mL(p)$ to be the inverse
function of $w=g(z)$, and $\tilde{\mL}(p)$ to be the inverse
function of $w=f(z)$  by replacing $w$ with $p$, then the
identities satisfied by the Faber polynomials of $f$ and $g$
\eqref{Faber10} say that
\begin{align}\label{Faber11}
\Phi_n &=(\mL^n)_{\geq 0} = \mL^n - \sum_{m=1}^{\infty}\frac{1}{m}
\frac{\pa^2 \F}{\pa t_n \pa t_m} \mL^{-m} =\frac{\pa^2 \F}{\pa
t_0\pa t_n} - \sum_{m=1}^{\infty} \frac{1}{m}\frac{\pa^2 \F}{\pa
t_{-m}\pa t_n}
\tilde{\mL}^m,\\
\Psi_n&= (\tilde{\mL}^{-n})_{\leq0} =-\frac{\pa^2 \F}{\pa t_0\pa
t_{-n}}-\sum_{m=1}^{\infty} \frac{1}{m} \frac{\pa^2 \F}{\pa t_m
\pa t_{-n}} \mL^{-m} = \tilde{\mL}^{-n} -\sum_{m=1}^{\infty}
\frac{1}{m} \frac{\pa^2 \F}{\pa t_{-m} \pa
t_{-n}}\tilde{\mL}^{m}.\nonumber
\end{align}
Comparing the coefficients of $p^0$ term in the last equation, we
have
\begin{align*}
(\tilde{\mL}^{-n})_{0}=-\frac{\pa^2 \F}{\pa t_0\pa t_{-n}}.
\end{align*}
Hence we can rewrite the last equation as
\begin{align}\label{Faber12}
(\tilde{\mL}^{-n})_{<0} =-\sum_{m=1}^{\infty} \frac{1}{m}
\frac{\pa^2 \F}{\pa t_m \pa t_{-n}} \mL^{-m} = \tilde{\mL}^{-n}
+\frac{\pa^2 \F}{\pa t_0\pa t_{-n}}-\sum_{m=1}^{\infty}
\frac{1}{m} \frac{\pa^2 \F}{\pa t_{-m} \pa t_{-n}}\tilde{\mL}^{m}.
\end{align}
Now from \eqref{identity4}, \eqref{Faber11} and \eqref{Faber12},
we have
\begin{align}\label{Lax1-3}
\frac{1}{\mathfrak{p}(\mL)} \frac{\pa \mathfrak{p}(\mL)}{\pa
t_n}\Bigr\vert_{\mL \fixed} &= -\sum_{m=1}^{\infty}
\frac{1}{m}\frac{\pa^3 \F}{\pa t_n \pa t_0
\pa t_m}\mL^{-m} = \frac{\pa (\mL^n)_{\geq 0}}{\pa t_0} \Bigr\vert_{\mL \fixed},\\
 \frac{1}{\mathfrak{p}(\mL)} \frac{\pa \mathfrak{p}(\mL)}{\pa
t_{-n}}\Bigr\vert_{\mL \fixed} &= -\sum_{m=1}^{\infty}
\frac{1}{m}\frac{\pa^3 \F}{\pa t_{-n} \pa t_0
\pa t_m}\mL^{-m}=\frac{\pa (\tilde{\mL}^{-n})_{<0}}{\pa t_{0}}\Bigr\vert_{\mL \fixed} ,\nonumber\\
\frac{1}{\tilde{\mathfrak{p}}(\tilde{\mL})} \frac{\pa
\tilde{\mathfrak{p}}(\tilde{\mL})}{\pa t_n}\Bigr\vert_{\tilde{\mL}
\fixed} &= \frac{\pa^3 \F}{\pa t_n \pa t_0^2} -\sum_{m=1}^{\infty}
\frac{1}{m}\frac{\pa^3 \F}{\pa t_n \pa
t_0 \pa t_{-m}}\tilde{\mL}^{m}=\frac{\pa (\mL^n)_{\geq 0}}{\pa t_0}\Bigr\vert_{\tilde{\mL} \fixed},\nonumber\\
\frac{1}{\tilde{\mathfrak{p}}(\tilde{\mL})} \frac{\pa
\tilde{\mathfrak{p}}(\tilde{\mL})}{\pa
t_{-n}}\Bigr\vert_{\tilde{\mL} \fixed} &= \frac{\pa^3 \F}{\pa
t_{-n} \pa t_0^2} -\sum_{m=1}^{\infty} \frac{1}{m}\frac{\pa^3
\F}{\pa t_{-n} \pa t_0 \pa t_{-m}}\tilde{\mL}^{m}=\frac{\pa
(\tilde{\mL}^{-n})_{<0}}{\pa t_{0}}\Bigr\vert_{\tilde{\mL}
\fixed}.\nonumber
\end{align}
On the other hand, since $\mathfrak{p}\circ \mL$ is the identity
function in $p$, by chain rule, we have
\begin{align}\label{equation9}
\frac{\pa \mathfrak{p}(\mL)}{\pa \mL}\frac{\pa \mL}{\pa t} +
\frac{\pa \mathfrak{p}(\mL)}{\pa t}\Bigr\vert_{\mL \fixed}=0,
\end{align}
and similarly for $\tilde{\mL}$. Here $t$ is any of the
independent variables. Hence
\begin{align*}
\frac{\pa \mL}{\pa t_n} &= - \frac{\pa \mL}{\pa p}\frac{\pa
\mathfrak{p}(\mL)}{\pa t_n}\Bigr\vert_{\mL \fixed}= -p \frac{\pa
\mL}{\pa p}\frac{\pa (\mL^n)_{\geq 0}}{\pa t_0} \Bigr\vert_{\mL
\fixed}\\
&=-p\frac{\pa \mL}{\pa p}\left(\frac{\pa (\mL^n)_{\geq 0}}{\pa
t_0}-\frac{\pa (\mL^n)_{\geq 0}}{\pa \mL}\frac{\pa \mL}{\pa
t_0}\right)= p\left(\frac{\pa (\mL^n)_{\geq 0}}{\pa p}\frac{\pa
\mL}{\pa t_0}-\frac{\pa \mL}{\pa p}\frac{\pa (\mL^n)_{\geq 0}}{\pa
t_0}\right)\\
&=\{ (\mL^n)_{\geq 0}, \mL\}_T,
\end{align*}
which is the first one of the Lax equations \eqref{Lax1}. The
other equations are derived in the same way from \eqref{Lax1-3}.
\end{proof}
\begin{remark}
The equations in \eqref{Lax1-3} are sometimes taken to be the
definition of dispersionless Toda hierarchy.
\end{remark}

There are two variants of the dToda hierarchy that we would like
to discuss here. Let $(\mL, \tilde{\mL})$ be a solution of the
dToda hierarchy and $\phi$ be the function defined by
\eqref{phi1}, then we can make a Miura-type transformation (Lemma
2.1.3 in \cite{TT1}) and define $\mL' = e^{-\ad \phi}\mL$,
$\tilde{\mL}'=e^{-\ad \phi}\tilde{\mL}$. Then they are of the form
\begin{align*}
\mL'(p)=& \tilde{u}_0p + \sum_{n=0}^{\infty} u'_{n+1}(t) p^{-n}=
\tilde{u}_0p + \sum_{n=0}^{\infty} u_{n+1}(t) (\tilde{u_0}p)^{-n},\\
\tilde{\mL'}^{-1}(p) =& p^{-1} + \sum_{n=0}^{\infty}
\tilde{u}'_{n+1}(t) p^n=p^{-1} + \sum_{n=0}^{\infty}
\tilde{u}_{n+1}(t) (\tilde{u_0}p)^n,
\end{align*}
and satisfy the Lax equations
\begin{align*}
\frac{\pa \mL'}{\pa t_n} =\{ ((\mL')^n)_{> 0}, \mL'\}_T,
\hspace{2cm}
\frac{\pa \mL'}{\pa t_{-n}} = \{((\tilde{\mL}')^{-n}))_{\leq0}, \mL'\}_T,\\
\frac{\pa \tilde{\mL'}}{\pa t_n} =\{ ((\mL')^n)_{> 0},
\tilde{\mL}'\}_T, \hspace{2cm} \frac{\pa \tilde{\mL'}}{\pa t_{-n}}
= \{((\tilde{\mL}')^{-n})_{\leq0}, \tilde{\mL}'\}_T.
\end{align*}
Now the dressing functions for $\mL'$ and $\tilde{\mL}'$ are
$\varphi'$ and $\tilde{\varphi}'$ respectively, where $e^{\ad
\varphi'} = e^{-\ad \phi}e^{\ad \varphi}$ and $e^{\ad
\tilde{\varphi}'} = e^{-\ad \phi}e^{\ad \tilde{\varphi}}$. The
Orlov-Schulman functions $\M'$ and $\tilde{\M}'$ are defined as in
\eqref{Orlov1}, and we find that
\begin{align*}
\M' = e^{-\ad \phi} \M &=  \sum_{n=1}^{\infty} nt_n (\mL')^n +
t_0+ \sum_{n=1}^{\infty} v^T_n
(\mL')^{-n},\\
\tilde{\M}'= e^{-\ad \phi} \tilde{\M} &=-\sum_{n=1}^{\infty}
nt_{-n} (\tilde{\mL}')^{-n} + t_0 - \sum_{n=1}^{\infty}
\tilde{v}^T_n (\tilde{\mL}')^{n}. \nonumber
\end{align*}
In particular, the functions $v^T_n$'s and $\tilde{v}_n^T$'s are
not changed. Hence the tau function is the same. If we look at
this transformation from the point of view of conformal maps, what
it amounts to is the pre-composition of $z=G(w)$ and $z=F(w)$ with
the linear map $w \mapsto \tilde{u_0}w$ ($\tilde{u}_0 =a_1$, the
leading coefficient of $f(z)$). Hence for the inverse function, we
have
\[
g'(z) = \frac{g(z)}{\tilde{u}_0}, \hspace{1cm}f'(z) =
\frac{f(z)}{\tilde{u}_0}.
\]
From these and the definition of the Grunsky coefficients, it is
quite obvious that the Grunsky coefficients are not changed. In
terms of the tau function, we now have
\begin{align*}
\mathfrak{p}(\mL') = g'(\mL') &= \mL'\exp\left(-\frac{\pa^2
\F}{\pa t_0^2}-\sum_{m=1}^{\infty}
\frac{1}{m}\frac{\pa^2 \F }{\pa t_0 \pa t_m}(\mL')^{-m}\right) ,\\
 \tilde{\mathfrak{p}}(\tilde{\mL}') =f'(\tilde{\mL}')
&=\tilde{\mL}' \exp\left( -\sum_{m=1}^{\infty} \frac{1}{m}
\frac{\pa^2\F}{\pa t_{-m}\pa t_0} (\tilde{\mL}')^m \right),
\end{align*}
and the dispersionless Hirota equations for $(\mL', \tilde{\mL}')$
assume the same form \eqref{bilinear1} .

Another variant of the dToda hierarchy is the one favored by
Wiegmann and Zabrodin in association with conformal maps \cite{WZ,
MWZ, KKMWZ}. Now the Miura type transformation is defined as
 $\mL' = e^{-\ad \phi/2}\mL$,
$\tilde{\mL}'=e^{-\ad \phi/2}\tilde{\mL}$. Then they are of the
form
\begin{align*}
\mL'(p)=& rp + \sum_{n=0}^{\infty} u'_{n+1}(t) p^{-n}=
 rp + \sum_{n=0}^{\infty} u_{n+1}(t) (rp)^{-n}\\
\tilde{\mL'}^{-1}(p) =& rp^{-1} + \sum_{n=0}^{\infty}
\tilde{u}'_{n+1}(t) p^n= rp^{-1} + \sum_{n=0}^{\infty}
\tilde{u}_{n+1}(t) (rp)^n,
\end{align*}
where $r$ is a square root of $\tilde{u}_0$. The Lax equations
become \footnote{Notice that the $\bar{t}_n$ in \cite{WZ} is the
complex conjugate of $t_n$. It amounts to $-t_{-n}$ here. }
\begin{align*}
\frac{\pa \mL'}{\pa t_n} &=\{ \mathcal{H}_n, \mL'\}_T,
\hspace{3cm}
\frac{\pa \mL'}{\pa t_{-n}} = \{\tilde{\mathcal{H}}_n, \mL'\}_T,\\
\frac{\pa \tilde{\mL'}}{\pa t_n} &=\{\mathcal{H}_n ,
\tilde{\mL}'\}_T, \hspace{3cm} \frac{\pa \tilde{\mL'}}{\pa t_{-n}}
=
\{\tilde{\mathcal{H}}_n, \tilde{\mL}'\}_T\\
\mathcal{H}_n &= ((\mL')^n)_{> 0}+ \frac{1}{2}((\mL')^n)_{0},
\hspace{1cm}\tilde{\mathcal{H}}_n= ((\tilde{\mL}')^{-n})_{<0} +
\frac{1}{2}((\tilde{\mL}')^{-n})_{0}.
\end{align*}
This version has the advantage that the roles of $\mL$ and $\mL'$
are symmetric. The same discussion above shows that the tau
function and dispersionless Hirota equations assume the same form
but now
\begin{align*}
\mathfrak{p}(\mL') = g'(\mL') &=
\mL'\exp\left(-\frac{1}{2}\frac{\pa^2 \F}{\pa
t_0^2}-\sum_{m=1}^{\infty}
\frac{1}{m}\frac{\pa^2 \F }{\pa t_0 \pa t_m}(\mL')^{-m}\right) ,\\
\tilde{ \mathfrak{p}}(\tilde{\mL}') =f'(\tilde{\mL}')
&=\tilde{\mL}' \exp\left( \frac{1}{2}\frac{\pa^2 \F}{\pa
t_0^2}-\sum_{m=1}^{\infty} \frac{1}{m} \frac{\pa^2\F}{\pa
t_{-m}\pa t_0} (\tilde{\mL}')^m \right).
\end{align*}

We can also view the dispersionless Hirota equations as a
consequence of the definition of the Grunsky coefficients for the
pair $(g , g_f)$. From this point of view and our discussion
above, we readily see that if $(\mL, \tilde{\mL})$ is a solution
to the first version of the dToda hierarchy, then the pair $(\mL',
\tilde{\mL}')$, where $\mL'(p) = \tilde{\mL}(1/p)^{-1}$,
$\tilde{\mL}'(p) = \mL(1/p)^{-1}$ is a solution to the second
version of the dToda hierarchy, if we redefine the independent
variables as $t_n' =t_{-n}$ and $t_{-n}' =t_n$.

\subsection{Dispersionless (coupled) modified KP hierarchy}
We define the dispersionless coupled modified KP hierarchy (dcmKP)
in \cite{T}. Here we are only interested in a special case. (In
the notation in \cite{T}, it is the case where $\mP =k$.)

\subsubsection{The hierarchy.}
The fundamental quantity is a formal power series
\begin{align*}
\mL = k + \sum_{n=0}^{\infty} u_{n+1}(t) k^{-n},
\end{align*}
with coefficients depending on the parameters $t= ( x, t_0, t_1
,t_2 \ldots)$. The Lax representation of the dcmKP in our special
case here is
\begin{align}\label{Lax2}
\frac{\pa \mL}{\pa t_n} = \{ (\mL^n)_{>0}, \mL\}, \hspace{1cm}
n\geq 1; \hspace{1cm} \frac{\pa \mL}{\pa t_0} = \frac{1}{k}
\frac{\pa \mL}{\pa x}.
\end{align}
Here $\{ \cdot, \cdot\}$ is the Poisson bracket defined as $\{f,
g\}= \frac{\pa f}{\pa k} \frac{\pa g}{\pa x} - \frac{\pa f}{\pa x}
\frac{\pa g}{\pa k} $. The $n=1$ flow of the hierarchy says that
\begin{align*}
\frac{\pa \mL}{\pa t_1}=\frac{\pa \mL}{\pa x}.
\end{align*}
Hence all the dependence on $t_1$ and $x$ appears in the form
$t_1+x$. There exists a dressing function $\varphi =
\sum_{n=0}^{\infty} \varphi_n(t) k^{-n}$ and a function $\phi(t)$
such that
\begin{align}
\mL &= e^{\ad \varphi} k, \hspace{3cm} \frac{\pa \phi}{\pa t_n} =
(\mL^n)_{0}, \hspace{0.2cm} n\geq 1,
\label{phi2}\\
\nabla_{t_n, \varphi} \varphi &= - (\mL^n)_{\leq 0},
\hspace{0.5cm}n\geq 1; \hspace{1cm} \nabla_{t_0, \varphi} \varphi
= \log k - \log \mL \nonumber.
\end{align}
Here $(\ad f)(g) =\{ f,g\}$. The Orlov-Schulman function is
defined as
\begin{align*}
\M =e^{\ad \varphi} \left( \sum_{n=1}^{\infty} nt_n k^{n-1} + x +
\frac{t_0}{k}\right) = \sum_{n=1}^{\infty} nt_n \mL^{n-1} + x+
\frac{t_0}{\mL} + \sum_{n=1}^{\infty} v_n \mL^{-n-1}.
\end{align*}
There exists a tau function $\tau$ such that \begin{align*}
\frac{\pa \log\tau}{\pa t_n} = v_n, \hspace{0.2cm} n\geq
1,\hspace{1cm} \frac{\pa \log \tau}{\pa t_0} = \phi.
\end{align*}
We define the free energy of the hierarchy by $\F= \log \tau$.

\subsubsection{Dispersionless Hirota equation.}
In terms of the tau function or the free energy, we have the
following identities:
\begin{align}\label{equation2}
\log k &= \log \mL -\sum_{m=1}^{\infty}\frac{1}{m} \frac{\pa^2\F}
{\pa t_m\pa t_0} \mL^{-m},\\
(\mL^n)_{> 0}&=\mL^n -\frac{\pa^2 \F}{\pa t_0 \pa
t_n}-\sum_{m=1}^{\infty} \frac{1}{m} \frac{\pa^2 \F}{\pa t_n \pa
t_m} \mL^{-m}. \nonumber
\end{align}
From the second equation, we also have
\begin{align}\label{equation3}
(\mL^n)_{\geq 0}&=\mL^n -\sum_{m=1}^{\infty} \frac{1}{m}
\frac{\pa^2 \F}{\pa t_n \pa t_m} \mL^{-m}.
\end{align}
Now we identify $k$ with $w$ and $\mL$ with $z$. The first
equation in \eqref{equation2} defines $w$ as a function of $z$,
which we denote by $g(z)$:
\begin{align}\label{equation4}
 g(z) =z \exp \left(-\sum_{m=1}^{\infty}\frac{1}{m}
\frac{\pa^2\F} {\pa t_m\pa t_0} z^{-m}\right).
\end{align}
Obviously, $g \in \Sigma$. The Faber polynomials $\Phi_n$ for $g$
are then identified with $(\mL^n)_{\geq 0}$. Comparing
\eqref{Faber3} to \eqref{equation3}, we find that the Grunsky
coefficients of $g(z)$ are related to the tau function or the free
energy by
\begin{align}\label{equation6}
b_{mn} = -\frac{1}{mn} \frac{\pa^2 \F}{\pa t_m \pa t_n}.
\end{align}
Together with \eqref{equation4}, we can rewrite the definition of
the Grunsky coefficients \eqref{Grunsky1} as
\begin{align}
&z_1\exp\left(-\sum_{m=1}^{\infty} \frac{1}{m}\frac{\pa^2 \F }{\pa
t_0 \pa t_m}z_1^{-m}\right)-z_2\exp\left(-\sum_{m=1}^{\infty}
\frac{1}{m}\frac{\pa^2 \F
}{\pa t_0 \pa t_m}z_2^{-m}\right)\label{bilinear2}\\
=&(z_1-z_2)\exp\left(\sum_{m,n=1}^{\infty} \frac{1}{mn}
\frac{\pa^2 \F}{\pa t_m \pa t_n}z_1^{-m}z_2^{-n}\right)\nonumber,
\end{align}
which is the dispersionless Hirota equation for this special case
of dcmKP hierarchy.

Conversely, we can characterize the tau function for dcmKP
hierarchy as:

\begin{proposition}\label{char2}
If $\F=\log \tau$ is a function of $t_n$, $n \geq 0$ that
satisfies the Hirota equation \eqref{bilinear2} and
\begin{align}\label{equation8} \frac{\pa^3 \log \tau }{\pa t_0^2 \pa t_1}=0,
\end{align}
 then $\tau$ is a tau function of a solution of the dcmKP
hierarchy. More explicitly, if we define $\mL(k)$ by formally
inverting the function $\mathsf{k}(\mL)$ defined by
\begin{align}\label{equation5}
\log \mathsf{k}(\mL) &= \log \mL-\sum_{m=1}^{\infty}
\frac{1}{m}\frac{\pa^2 \F }{\pa t_0 \pa t_m}\mL^{-m},
\end{align}
and replacing $t_1$ by $t_1+x$, then $\mL$ satisfies the Lax
equations \eqref{Lax2} for dcmKP hierarchy.
\end{proposition}
\begin{proof}
The proof is almost the same as the case of dToda. Identifying
$\mL$ with $z$, we define the function $g(z) \in \Sigma$ by
\begin{align*}
\log g(z) = \log z - \sum_{m=1}^{\infty} \frac{1}{m}\frac{\pa^2 \F
}{\pa t_0 \pa t_m}z^{-m}.
\end{align*}
The dispersionless Hirota equation \eqref{bilinear2} then read as
\begin{align*}
\log \frac{g(z_1) - g(z_2)}{z_1-z_2} = \sum_{m, n =1}^{\infty}
\frac{1}{mn} \frac{\pa^2 \F}{\pa t_m\pa t_n} z_1^{-m} z_2^{-n}.
\end{align*}
Comparing with the definition of the Grunsky coefficients of $g$
given by equation \eqref{Grunsky1}, we find that the relation
between the Grunsky coefficients and free energy is given by
\eqref{equation6} . Let $z=G(w)$ be the inverse of $g(z)$. Then
$G(w)$ is the function $\mL(k)$ defined by \eqref{equation5} if we
identify $k$ with $w$. The Faber polynomials of $g(z)$ is then
identified with $(\mL(w))_{\geq 0}$. We can then rewrite the
identity satisfied by the Faber polynomials \eqref{Faber3} in
terms of the free energy by
\begin{align*}
(\mL^n)_{\geq 0} = \mL^n -\sum_{m=1}^{\infty} \frac{1}{m}
\frac{\pa^2 \F}{\pa t_m \pa t_n} \mL^{-m}.
\end{align*}
From \eqref{equation5}, we also have
\begin{align*}
\frac{\pa^2 \F}{\pa t_n \pa t_0} = \Res \mL^n d \log k=(\mL^n)_0.
\end{align*}
Hence
\begin{align}\label{equation7}
(\mL^n)_{>0} = \mL^n -\frac{\pa^2 \F}{\pa t_n \pa
t_0}-\sum_{m=1}^{\infty} \frac{1}{m} \frac{\pa^2 \F}{\pa t_m \pa
t_n} \mL^{-m}.
\end{align}
From \eqref{equation5} and the $n=1$ case in \eqref{equation7}, we
have
\begin{align*}
\frac{1}{\mathsf{k}} \frac{\pa \mathsf{k}}{\pa t_1}
\Bigr\vert_{\mL \fixed} = -\sum_{m=1}^{\infty} \frac{1}{m}
\frac{\pa^3 \F}{\pa t_m \pa t_0 \pa t_1} = \frac{\pa
\mathsf{k}}{\pa t_0} \Bigr\vert_{\mL \fixed}.
\end{align*}
We have used $(\mL)_{>0} =k$ and \eqref{equation8}. Using the
identity \eqref{equation9} (with $p$ replaced by $k$), it follows
that
\begin{align*}
\frac{\pa \mL}{\pa x} = \frac{\pa \mL}{\pa t_1} = k\frac{\pa
\mL}{\pa t_0}.
\end{align*}
This gives the second equation in the Lax equations \eqref{Lax2}.
Using the $n=1$ case in \eqref{equation7} and  $(\mL)_{>0} =k$
again, we have
\begin{align*}
\frac{\pa k}{\pa t_n}\Bigr\vert_{\mL \fixed} =-\frac{\pa^2 \F}{\pa
t_1 \pa t_n \pa t_0}-\sum_{m=1}^{\infty} \frac{1}{m} \frac{\pa^2
\F}{\pa t_m \pa t_n \pa t_1} \mL^{-m} =\frac{\pa (\mL^n)_{>0}}{\pa
t_1}\Bigr\vert_{\mL \fixed},
\end{align*}
which by \eqref{equation9} is equivalent to
\begin{align*}
 \frac{\pa \mL}{\pa t_n} =& - \frac{\pa \mL}{\pa k}\frac{\pa (\mL^n)_{>0}}{\pa
t_1} \Bigr\vert_{\mL \fixed}= -\frac{\pa \mL}{\pa
k}\left(\frac{\pa (\mL^n)_{>0}}{\pa x} -\frac{\pa
(\mL^n)_{>0}}{\pa \mL}\frac{\pa \mL}{\pa x}\right)\\
=&\{ (\mL^n)_{>0}, \mL\}.
\end{align*}
\end{proof}

Compare the dispersionless Hirota equations for dToda and dcmKP
hierarchies, we immediately have
\begin{corollary}\label{cor1}
If $(\mL, \tilde{\mL})$ is a solution to the dToda hierarchy
\eqref{Lax1}, and $\frac{\pa (\mL)_0}{\pa t_0} =0$, then $\mL$ is
a solution to the dcmKP hierarchy \eqref{Lax2}, when we replace
$t_1$ by $t_1 + x$ and regard the $t_{-n}$'s, $n\geq 1$ as
parameters. The tau function for the dToda hierarchy is the
 tau function for the corresponding dcmKP hierarchy.
\end{corollary}
\begin{proof}
In \cite{T}, we proved this proposition by comparing the Lax
equations. Here we just notice that the first equation in the
Hirota equations for dToda \eqref{bilinear1} is identical with the
Hirota equation for dcmKP \eqref{bilinear2}. The result follows
from the proposition above.

\end{proof}

\subsection{Dispersionless KP hierarchy}.
This is the most well known case. The dispersionless Hirota
equation for dKP hierarchy was first derived as the
quasi-classical limit of the differential Fay identity by Takasaki
and Takebe in \cite{TT1}, see also the work of Carroll and Kodama
\cite{CK}. Here we derive the Hirota equation along the same line
as we do for dToda and dcmKP hierarchies.
\subsubsection{The hierarchy.}
The fundamental quantity in dKP hierarchy is a formal power series
\begin{align*}
\mL = k+ \sum_{n=1}^{\infty} u_{n+1}(t) k^{-n}
\end{align*}
with coefficients depending on the independent variables $t=(x,
t_1, t_2 , \ldots)$. The Lax equation is
\begin{align}\label{Lax3}
\frac{\pa \mL}{\pa t_n} = \{ (\mL^n)_{\geq 0}, \mL\}.
\end{align}
Here the Poisson bracket is the same as in the dcmKP hierarchy.
Also, the first equation $n=1$ of the hierarchy says that the
dependence on $t_1$ and $x$ appears in the combination $t_1+x$.
There exists a dressing function $\varphi = \sum_{n=1}^{\infty}
\varphi_n k^{-n}$ such that
\begin{align*}
\mL = e^{\ad \varphi} k, \hspace{2cm} \nabla_{t_n, \varphi}
\varphi = -(\mL^n)_{<0}.
\end{align*}
With this dressing function, the Orlov-Schulman function is
defined by
\begin{align*}
\M = e^{\ad \varphi} \left( \sum_{n=1}^{\infty} nt_n k^{n-1}
+x\right) = \sum_{n=1}^{\infty} nt_n \mL^{n-1} +x
+\sum_{n=1}^{\infty} v_n \mL^{-n-1}.
\end{align*}
There exists a tau function $\tau$ such that
\begin{align*}
\frac{\pa \log \tau}{\pa t_n} = v_n.
\end{align*}
We define the free energy by $\F = \log \tau$.
\subsubsection{Dispersionless Hirota equation}
In terms of the tau function or the free energy, we have the
following identities:
\begin{align}\label{eqn2}
(\mL^n)_{\geq 0} = \mL^n -\sum_{m=1}^{\infty} \frac{1}{m}
\frac{\pa^2 \F}{\pa t_m \pa t_n} \mL^{-m}.
\end{align}
Now we identify $k$ with $w$ and $\mL$ with $z$, the function
$z=G(w)$ is defined to be $\mL(k)$, and the function $w=g(z) \in
\Sigma_0$ the inverse of $G(w)$. Then the Faber polynomials
$\Phi_n(w)$ of $g$ are identified with $(\mL^n(w))_{\geq 0}$.
Define the Grunsky coefficients of $g$ by equation
\eqref{Grunsky1}, then comparing \eqref{Faber3} with \eqref{eqn2},
we find the relation between the Grunsky coefficients and the free
energy is given by
\begin{align}\label{eqn3}
b_{mn} = -\frac{1}{mn} \frac{\pa^2 \F}{\pa t_m \pa t_n}.
\end{align}
From the $n=1$ case of equation \eqref{eqn2} and the fact that
$(\mL)_{\geq 0}=k$, we have
\begin{align*}
g(z) = z -\sum_{m=1}^{\infty} \frac{1}{m} \frac{\pa^2 \F}{\pa t_m
\pa t_1} z^{-m}.
\end{align*}
Using this, and \eqref{eqn3}, we can rewrite the definition of the
Grunsky coefficients \eqref{Grunsky1} in terms of the free energy:
\begin{align}\label{bilinear3}
1- \frac{1}{z_1-z_2}\sum_{m=1}^{\infty} \frac{z_1^{-m}- z_2^{-m}
}{m} \frac{\pa^2 \F}{\pa t_m \pa t_1}= \exp\left(
\sum_{m,n=1}^{\infty} \frac{1}{mn} \frac{\pa^2\F}{\pa t_m\pa t_n}
z_1^{-m} z_2^{-n}\right),
\end{align}
which is the dispersionless Hirota equation for dKP.

Conversely, we can characterize the tau function for the dKP
hierarchy as follows:
\begin{proposition}\label{char3}
If $\F=\log \tau$ is a function of $t_n$, $n \geq 1$ that
satisfies the Hirota equation \eqref{bilinear3},
 then $\tau$ is a tau function of a solution of the dKP
hierarchy. More explicitly, if we define $\mL(k)$ by formally
inverting the function $\mathsf{k}(\mL)$ defined by
\begin{align}\label{eqn4}
\mathsf{k}(\mL) &=  \mL-\sum_{m=1}^{\infty} \frac{1}{m}\frac{\pa^2
\F }{\pa t_1 \pa t_m}\mL^{-m},
\end{align}
and replacing $t_1$ by $t_1+x$, then $\mL$ satisfies the Lax
equations \eqref{Lax3} for dKP.
\end{proposition}
\begin{proof}
The proof follows the same idea as Propositions \ref{char1} and
\ref{char2}. We define the function $g \in \Sigma_0$ by
identifying $\mL$ with $z$ in \eqref{eqn4}, namely
\begin{align*}
g(z) = z -\sum_{m=1}^{\infty} \frac{1}{m}\frac{\pa^2 \F }{\pa t_1
\pa t_m}z^{-m}.
\end{align*}
Hence the dispersionless Hirota equation \eqref{bilinear3} says
that
\begin{align}\label{eqn5}
\log \frac{g(z_1) - g(z_2)}{z_1 -z_2} = \sum_{m,n=1}^{\infty}
\frac{1}{mn}\frac{\pa^2 \F}{\pa t_m \pa t_n} z_1^{-m} z_2^{-n}.
\end{align}
Compare with the definition of the Grunsky coeffcients $b_{mn}$
\eqref{Grunsky1} of $g$, we find that the $b_{mn}$ can be
expressed in terms of the free energy by \eqref{eqn3}. Now define
$z=G(w)$ to be the formal inverse of $w=g(z)$. In other words,
$G(w)$ is $\mL(k)$ if we identify $k$ with $w$. Then the Faber
polynomials $\Phi_n(w)$ of $g$ are identified with
$(\mL^n(k))_{\geq 0}$. Hence the identities satisfied by the Faber
polynomials \eqref{Faber3} can be rewritten as
\begin{align}\label{eqn6}
(\mL^n)_{\geq 0} = \mL^n -\sum_{m=1}^{\infty} \frac{1}{m}
\frac{\pa^2\F}{\pa t_m \pa t_n}\mL^{-m}.
\end{align}
Fixing $\mL$, differentiate the $n=1$ case with respect to $t_n$
and compare to the result when we differentiate the $n$ case with
respect to $t_1$, we have
\begin{align*}
\frac{\pa \mathsf{k}}{\pa t_n} \Bigr\vert_{\mL \fixed} =
-\sum_{m=1}^{\infty} \frac{1}{m}\frac{\pa^3 \F}{\pa t_m \pa t_1
\pa t_n} \mL^{-m} = \frac{\pa (\mL^n)_{\geq 0} }{\pa t_1}
\Bigr\vert_{\mL \fixed}.
\end{align*}
The same argument as in Proposition \ref{char2} gives the Lax
equation \eqref{Lax3} of dKP.

\end{proof}

From this characterization of the tau functions, we can also see
that a solution of the dcmKP hierarchy will give rise to a
solution of the dKP hierarchy.
\begin{corollary}\label{cor2}
If
\[\mL= k+ \sum_{n=0}^{\infty} u_{n+1}k^{-n}
\]
 is a solution of the dcmKP hierarchy, then
 \[
 \mL' = k+
\sum_{n=1}^{\infty} u'_{n+1}k^{-n} =k + \sum_{n=1}^{\infty}
u_{n+1}(k-u_1)^{-n} \]
 is a solution of the dKP hierarchy, where we regard $t_0$ as a
 constant. Moreover, $\mL$ and $\mL'$ have the same tau function.
\end{corollary}
\begin{proof}
In \cite{CT, T}, this was proved via a Miura map $\mL' = e^{\ad
\phi}\mL$ to relate $\mL$ and $\mL'$, where $\phi$ is defined in
\eqref{phi2}. Here we give a proof in terms of tau functions.

Since $\mL=k+ \sum_{n=0}^{\infty} u_{n+1}k^{-n}$ is a solution of
the dcmKP, the free energy of $\F$ satisfies the dispersionless
Hirota equation \eqref{bilinear2}. Taking the limit $z_2
\rightarrow \infty$ in \eqref{bilinear2}, we obtain the relation
\begin{align*}
z_1\exp\left(-\sum_{m=1}^{\infty} \frac{1}{m}\frac{\pa^2 \F }{\pa
t_0 \pa t_m}z_1^{-m}\right)+ \frac{\pa^2 \F }{\pa t_0 \pa t_1}
=z_1-\sum_{m=1}^{\infty} \frac{1}{m} \frac{\pa^2 \F}{\pa t_m \pa
t_1}z_1^{-m},
\end{align*}
or
\begin{align*}
z\exp\left(-\sum_{m=1}^{\infty} \frac{1}{m}\frac{\pa^2 \F }{\pa
t_0 \pa t_m}z^{-m}\right) =z- \frac{\pa^2 \F }{\pa t_0 \pa
t_1}-\sum_{m=1}^{\infty} \frac{1}{m} \frac{\pa^2 \F}{\pa t_m \pa
t_1}z^{-m}.
\end{align*}
Substituting this into the Hirota equation \eqref{bilinear2}
again, we obtain
\begin{align*}
z_1 - z_2 -\sum_{m=1}^{\infty} \frac{z_1^m -z_2^m}{m} \frac{\pa^2
\F}{\pa t_m \pa t_1} = (z_1-z_2)\exp\left(\sum_{m,n=1}^{\infty}
\frac{1}{mn} \frac{\pa^2 \F}{\pa t_m \pa
t_n}z_1^{-m}z_2^{-n}\right),
\end{align*}
which is equivalent to the dispersionless Hirota equation for dKP
\eqref{bilinear3}. Hence, from Proposition \ref{char3} above, the
function $\mL'$ defined by inverting
\begin{align*}
\mathsf{k}(\mL') = \mL' - \sum_{m=1}^{\infty} \frac{1}{m}
\frac{\pa^2 \F}{\pa t_m \pa t_1}(\mL')^{-m}
\end{align*}
is a solution of the dKP hierarchy. Now since the inverse
$\mathsf{k}$ of $\mL(k)$ satisfies
\begin{align*}
\mathsf{k}(\mL) =\mL - \frac{\pa^2\F}{\pa t_0 \pa t_1}
-\sum_{m=1}^{\infty} \frac{1}{m} \frac{\pa^2 \F}{\pa t_m \pa
t_1}\mL^{-m},
\end{align*}
and $\frac{\pa^2\F}{\pa t_0 \pa t_1}=u_1$, we find that
\begin{align*}
\mL' = k + \sum_{n=1}^{\infty} u_{n+1}(k-u_1)^{-n} .
\end{align*}
\end{proof}

\begin{remark}
The proof of this corollary also shows that the dispersionless
Hirota equation for dcmKP implies the dispersionless Hirota
equation for dKP.
\end{remark}


\section{Concluding remarks}
We extend the definition of Grunsky coefficients and Faber
polynomials formally to the space of formal power series. We have
shown that the tau function or free energy of dToda, dcmKP and dKP
hierarchies are closely related to Grunsky coefficients. We
rederive the dispersionless Hirota equations and prove that they
uniquely characterize the tau functions associated to a solution
of the hierarchies. This might be helpful in classifying the
solutions of the hierarchies. We also establish a link between
these three dispersionless hierarchies. The transformation that
relate the three versions of the dToda hierarchies and the Miura
map that transform a solution of the dcmKP hierarchy to a solution
of the dKP hierarchy are just the linear maps that relate the
three spaces of formal power series we discuss in Section 2.

Given any formal power series $f\in \tilde{S}$, $g \in \Sigma$, if
we define $\F_{mn}$'s as
\begin{align*}
\F_{m,n} &= -|mn| b_{m,n}, \hspace{1cm} m\neq 0, n\neq 0,\\
 \F_{m,0}
&= \F_{0,m} = |m| b_{m,0}, \hspace{1cm} m \neq 0,
\hspace{2cm}\F_{0,0} = - b_{0,0},
\end{align*}
where $b_{m,n}$'s are the Grunsky coefficients associated to the
pair $(f, g)$, then $\F_{m,n}$'s satisfy the dispersionless Hirota
equations \eqref{bilinear1}, \eqref{bilinear2} and
\eqref{bilinear3} if we replace $\frac{\pa^2 \F}{\pa t_m \pa t_n}$
by $\F_{m,n}$. In \cite{BS}, Sorin and Bonora proved that the
Neumann coefficients that appear in string field theory satisfy
the dispersionless Hirota bilinear equations. By definition, the
Neumann coefficients coincide with the Grunsky coefficients
$b_{m,n}$ defined above. This explain their results.

However, it is still an open question to find a function $\F$ such
that
\begin{align*}
\frac{\pa^2 \F}{\pa t_m \pa t_n} = \F_{m,n}.
\end{align*}
In \cite{WZ}, Wiegmann and Zabrodin provided a solution to this
problem (see also \cite{KKMWZ, MWZ, Z}) when $G$, the inverse
function of $g$ is an analytic function that maps the outer disc
$\{|z|>1\}$ to the exterior of an analytic curve, and $f(z) =
1/\bar{g}(z^{-1})$. It will be interesting to solve the general
problem.

\vspace{0.3cm}

 \noindent \textbf{Acknowledgments.} I am grateful to L. A. Takhtajan
 for his helpful comments.
  This work is partially supported by
NSC grant NSC 91-2115-M-009-017.

\bibliographystyle{amsplain}
\bibliography{hirota}

\end{document}